\documentclass[electronic]{vgtc}             %

\makeatletter
\def\@fnsymbol#1{\ensuremath{\ifcase#1\or \medsquare\or \medcircle\or
   \medtriangledown\or \medtriangleup\or \|\or **\or \dagger\dagger
   \or \ddagger\ddagger \else\@ctrerr\fi}}
\makeatother

\ifpdf%
  \pdfoutput=1\relax                   %
  \usepackage{graphicx}                %
  \DeclareGraphicsExtensions{.pdf,.png,.jpg,.jpeg} %
\else%
  \usepackage{graphicx}                %
  \DeclareGraphicsExtensions{.eps}     %
\fi%

\graphicspath{{figures/}{pictures/}{images/}{./}} %

\usepackage{microtype}                 %
\PassOptionsToPackage{warn}{textcomp}  %
\usepackage{textcomp}                  %
\usepackage{mathptmx}                  %
\usepackage{times}                     %
\usepackage{cite}                      %
\usepackage{tabu}                      %
\usepackage{booktabs}                  %
\onlineid{0}

\vgtccategory{Research}

\vgtcinsertpkg

\title{\tool{}: A Practical Method for Creating Notebook-Ready Visual Analytics}

\author{
  Zijie J. Wang\thanks{e-mail: \href{mailto:jayw@gatech.edu}{jayw@gatech.edu}} %
  \and David Munechika\thanks{e-mail: \href{mailto:david.munechika@gatech.edu}{david.munechika@gatech.edu}} %
  \and Seongmin Lee\thanks{e-mail: \href{mailto:seongmin@gatech.edu}{seongmin@gatech.edu}} %
  \and Duen Horng Chau\thanks{e-mail: \href{mailto:polo@gatech.edu}{polo@gatech.edu}}
}
\affiliation{\vspace{-10pt}\scriptsize{Georgia Institute of Technology}}

\teaser{
  \centering
  \vspace{-8pt}
  \setlength{\belowcaptionskip}{0pt}
  \setlength{\abovecaptionskip}{0pt}
  \includegraphics[width=1\textwidth]{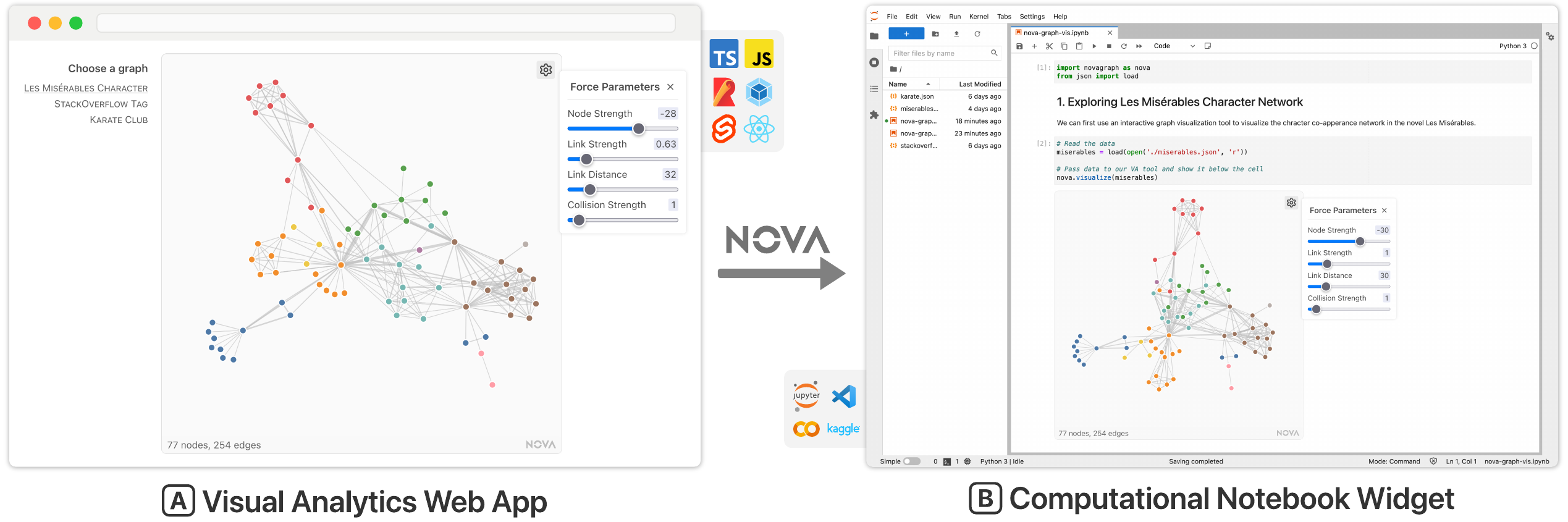}
  \caption{
    \tool{} allows visualization researchers and practitioners to quickly adapt existing web-based visual analytics tools to support computational notebooks.
    \textbf{(A)} \textbf{Web apps} are a popular medium for developing interactive visualization systems that users can access in their web browsers.
    \tool{} converts web apps developed with diverse web technologies, such as programming languages and frameworks, to \textbf{(B)} \textbf{notebook widgets} that end-users can easily install and use in different notebook environments.
  }
  \label{fig:teaser}
  \setlength{\belowcaptionskip}{0pt}
  \setlength{\abovecaptionskip}{10pt}
}

\abstract{
  How can we develop visual analytics (VA) tools that can be easily adopted?
  Visualization researchers have developed a large number of web-based VA tools to help data scientists in a wide range of tasks.
  However, adopting these standalone systems can be challenging, as they require data scientists to create new workflows to streamline the VA processes.
  Recent surveys suggest computational notebooks have been dominating data scientists' analytical workflows, as these notebooks seamlessly combine text, code, and visualization, allowing users to rapidly iterate code experiments.
  To help visualization researchers develop VA tools that can be easily integrated into existing data science workflows, we present \tool{}, a simple and flexible method to adapt web-based VA systems for notebooks.
  We provide detailed examples of using this method with diverse web development technologies and different types of computational notebooks.
  Deployed application examples highlight that \tool{} is easy to adopt, and data scientists appreciate in-notebook VA.
  \tool{} is available at \link{https://github.com/poloclub/nova}.
}

\CCScatlist{
  \CCScatTwelve{Human-centered computing}{Visu\-al\-iza\-tion}{}{}
}

\usepackage{enumitem}
\usepackage{tabu}
\usepackage{rotating}
\usepackage{tikz}
\usepackage{capt-of,etoolbox}
\usepackage{microtype}
\usepackage{bm}
\usepackage{soul}
\usepackage{wrapfig}
\usepackage{graphbox}
\usepackage{tcolorbox}
\usepackage{amssymb}
\usepackage{amsmath}
\usepackage{cleveref}
\usepackage{mathtools}
\usepackage{balance}
\usepackage[varqu]{zi4}
\usepackage[bb=boondox]{mathalfa}
\usepackage[utf8x]{inputenc}
\usepackage{units}
\usepackage{setspace}
\usepackage{gensymb}
\usepackage{helvet}
\usepackage{fdsymbol}
\usepackage{printlen} %
\usepackage[numbers]{natbib}

\usepackage{eqnarray}
\usepackage{bbm}

\definecolor{redOV}{RGB}{255, 235, 238}
\definecolor{redI}{RGB}{255, 205, 210}
\definecolor{redII}{RGB}{239, 154, 154}
\definecolor{redIII}{RGB}{229, 115, 115}
\definecolor{redIV}{RGB}{239, 83, 80}
\definecolor{redV}{RGB}{244, 67, 54}
\definecolor{redVI}{RGB}{229, 57, 53}
\definecolor{redVII}{RGB}{211, 47, 47}
\definecolor{redVIII}{RGB}{198, 40, 40}
\definecolor{redIX}{RGB}{183, 28, 28}
\definecolor{redAI}{RGB}{255, 138, 128}
\definecolor{redAII}{RGB}{255, 82, 82}
\definecolor{redAIV}{RGB}{255, 23, 68}
\definecolor{redAVII}{RGB}{213, 0, 0}

\definecolor{pinkOV}{RGB}{252, 228, 236}
\definecolor{pinkI}{RGB}{248, 187, 208}
\definecolor{pinkII}{RGB}{244, 143, 177}
\definecolor{pinkIII}{RGB}{240, 98, 146}
\definecolor{pinkIV}{RGB}{236, 64, 122}
\definecolor{pinkV}{RGB}{233, 30, 99}
\definecolor{pinkVI}{RGB}{216, 27, 96}
\definecolor{pinkVII}{RGB}{194, 24, 91}
\definecolor{pinkVIII}{RGB}{173, 20, 87}
\definecolor{pinkIX}{RGB}{136, 14, 79}
\definecolor{pinkAI}{RGB}{255, 128, 171}
\definecolor{pinkAII}{RGB}{255, 64, 129}
\definecolor{pinkAIV}{RGB}{245, 0, 87}
\definecolor{pinkAVII}{RGB}{197, 17, 98}

\definecolor{purpleOV}{RGB}{243, 229, 245}
\definecolor{purpleI}{RGB}{225, 190, 231}
\definecolor{purpleII}{RGB}{206, 147, 216}
\definecolor{purpleIII}{RGB}{186, 104, 200}
\definecolor{purpleIV}{RGB}{171, 71, 188}
\definecolor{purpleV}{RGB}{156, 39, 176}
\definecolor{purpleVI}{RGB}{142, 36, 170}
\definecolor{purpleVII}{RGB}{123, 31, 162}
\definecolor{purpleVIII}{RGB}{106, 27, 154}
\definecolor{purpleIX}{RGB}{74, 20, 140}
\definecolor{purpleAI}{RGB}{234, 128, 252}
\definecolor{purpleAII}{RGB}{224, 64, 251}
\definecolor{purpleAIV}{RGB}{213, 0, 249}
\definecolor{purpleAVII}{RGB}{170, 0, 255}

\definecolor{deeppurpleOV}{RGB}{237, 231, 246}
\definecolor{deeppurpleI}{RGB}{209, 196, 233}
\definecolor{deeppurpleII}{RGB}{179, 157, 219}
\definecolor{deeppurpleIII}{RGB}{149, 117, 205}
\definecolor{deeppurpleIV}{RGB}{126, 87, 194}
\definecolor{deeppurpleV}{RGB}{103, 58, 183}
\definecolor{deeppurpleVI}{RGB}{94, 53, 177}
\definecolor{deeppurpleVII}{RGB}{81, 45, 168}
\definecolor{deeppurpleVIII}{RGB}{69, 39, 160}
\definecolor{deeppurpleIX}{RGB}{49, 27, 146}
\definecolor{deeppurpleAI}{RGB}{179, 136, 255}
\definecolor{deeppurpleAII}{RGB}{124, 77, 255}
\definecolor{deeppurpleAIV}{RGB}{101, 31, 255}
\definecolor{deeppurpleAVII}{RGB}{98, 0, 234}

\definecolor{indigoOV}{RGB}{232, 234, 246}
\definecolor{indigoI}{RGB}{197, 202, 233}
\definecolor{indigoII}{RGB}{159, 168, 218}
\definecolor{indigoIII}{RGB}{121, 134, 203}
\definecolor{indigoIV}{RGB}{92, 107, 192}
\definecolor{indigoV}{RGB}{63, 81, 181}
\definecolor{indigoVI}{RGB}{57, 73, 171}
\definecolor{indigoVII}{RGB}{48, 63, 159}
\definecolor{indigoVIII}{RGB}{40, 53, 147}
\definecolor{indigoIX}{RGB}{26, 35, 126}
\definecolor{indigoAI}{RGB}{140, 158, 255}
\definecolor{indigoAII}{RGB}{83, 109, 254}
\definecolor{indigoAIV}{RGB}{61, 90, 254}
\definecolor{indigoAVII}{RGB}{48, 79, 254}

\definecolor{blueOV}{RGB}{227, 242, 253}
\definecolor{blueI}{RGB}{187, 222, 251}
\definecolor{blueII}{RGB}{144, 202, 249}
\definecolor{blueIII}{RGB}{100, 181, 246}
\definecolor{blueIV}{RGB}{66, 165, 245}
\definecolor{blueV}{RGB}{33, 150, 243}
\definecolor{blueVI}{RGB}{30, 136, 229}
\definecolor{blueVII}{RGB}{25, 118, 210}
\definecolor{blueVIII}{RGB}{21, 101, 192}
\definecolor{blueIX}{RGB}{13, 71, 161}
\definecolor{blueAI}{RGB}{130, 177, 255}
\definecolor{blueAII}{RGB}{68, 138, 255}
\definecolor{blueAIV}{RGB}{41, 121, 255}
\definecolor{blueAVII}{RGB}{41, 98, 255}

\definecolor{lightblueOV}{RGB}{225, 245, 254}
\definecolor{lightblueI}{RGB}{179, 229, 252}
\definecolor{lightblueII}{RGB}{129, 212, 250}
\definecolor{lightblueIII}{RGB}{79, 195, 247}
\definecolor{lightblueIV}{RGB}{41, 182, 246}
\definecolor{lightblueV}{RGB}{3, 169, 244}
\definecolor{lightblueVI}{RGB}{3, 155, 229}
\definecolor{lightblueVII}{RGB}{2, 136, 209}
\definecolor{lightblueVIII}{RGB}{2, 119, 189}
\definecolor{lightblueIX}{RGB}{1, 87, 155}
\definecolor{lightblueAI}{RGB}{128, 216, 255}
\definecolor{lightblueAII}{RGB}{64, 196, 255}
\definecolor{lightblueAIV}{RGB}{0, 176, 255}
\definecolor{lightblueAVII}{RGB}{0, 145, 234}

\definecolor{cyanOV}{RGB}{224, 247, 250}
\definecolor{cyanI}{RGB}{178, 235, 242}
\definecolor{cyanII}{RGB}{128, 222, 234}
\definecolor{cyanIII}{RGB}{77, 208, 225}
\definecolor{cyanIV}{RGB}{38, 198, 218}
\definecolor{cyanV}{RGB}{0, 188, 212}
\definecolor{cyanVI}{RGB}{0, 172, 193}
\definecolor{cyanVII}{RGB}{0, 151, 167}
\definecolor{cyanVIII}{RGB}{0, 131, 143}
\definecolor{cyanIX}{RGB}{0, 96, 100}
\definecolor{cyanAI}{RGB}{132, 255, 255}
\definecolor{cyanAII}{RGB}{24, 255, 255}
\definecolor{cyanAIV}{RGB}{0, 229, 255}
\definecolor{cyanAVII}{RGB}{0, 184, 212}

\definecolor{tealOV}{RGB}{224, 242, 241}
\definecolor{tealI}{RGB}{178, 223, 219}
\definecolor{tealII}{RGB}{128, 203, 196}
\definecolor{tealIII}{RGB}{77, 182, 172}
\definecolor{tealIV}{RGB}{38, 166, 154}
\definecolor{tealV}{RGB}{0, 150, 136}
\definecolor{tealVI}{RGB}{0, 137, 123}
\definecolor{tealVII}{RGB}{0, 121, 107}
\definecolor{tealVIII}{RGB}{0, 105, 92}
\definecolor{tealIX}{RGB}{0, 77, 64}
\definecolor{tealAI}{RGB}{167, 255, 235}
\definecolor{tealAII}{RGB}{100, 255, 218}
\definecolor{tealAIV}{RGB}{29, 233, 182}
\definecolor{tealAVII}{RGB}{0, 191, 165}

\definecolor{greenOV}{RGB}{232, 245, 233}
\definecolor{greenI}{RGB}{200, 230, 201}
\definecolor{greenII}{RGB}{165, 214, 167}
\definecolor{greenIII}{RGB}{129, 199, 132}
\definecolor{greenIV}{RGB}{102, 187, 106}
\definecolor{greenV}{RGB}{76, 175, 80}
\definecolor{greenVI}{RGB}{67, 160, 71}
\definecolor{greenVII}{RGB}{56, 142, 60}
\definecolor{greenVIII}{RGB}{46, 125, 50}
\definecolor{greenIX}{RGB}{27, 94, 32}
\definecolor{greenAI}{RGB}{185, 246, 202}
\definecolor{greenAII}{RGB}{105, 240, 174}
\definecolor{greenAIV}{RGB}{0, 230, 118}
\definecolor{greenAVII}{RGB}{0, 200, 83}

\definecolor{lightgreenOV}{RGB}{241, 248, 233}
\definecolor{lightgreenI}{RGB}{220, 237, 200}
\definecolor{lightgreenII}{RGB}{197, 225, 165}
\definecolor{lightgreenIII}{RGB}{174, 213, 129}
\definecolor{lightgreenIV}{RGB}{156, 204, 101}
\definecolor{lightgreenV}{RGB}{139, 195, 74}
\definecolor{lightgreenVI}{RGB}{124, 179, 66}
\definecolor{lightgreenVII}{RGB}{104, 159, 56}
\definecolor{lightgreenVIII}{RGB}{85, 139, 47}
\definecolor{lightgreenIX}{RGB}{51, 105, 30}
\definecolor{lightgreenAI}{RGB}{204, 255, 144}
\definecolor{lightgreenAII}{RGB}{178, 255, 89}
\definecolor{lightgreenAIV}{RGB}{118, 255, 3}
\definecolor{lightgreenAVII}{RGB}{100, 221, 23}

\definecolor{limeOV}{RGB}{249, 251, 231}
\definecolor{limeI}{RGB}{240, 244, 195}
\definecolor{limeII}{RGB}{230, 238, 156}
\definecolor{limeIII}{RGB}{220, 231, 117}
\definecolor{limeIV}{RGB}{212, 225, 87}
\definecolor{limeV}{RGB}{205, 220, 57}
\definecolor{limeVI}{RGB}{192, 202, 51}
\definecolor{limeVII}{RGB}{175, 180, 43}
\definecolor{limeVIII}{RGB}{158, 157, 36}
\definecolor{limeIX}{RGB}{130, 119, 23}
\definecolor{limeAI}{RGB}{244, 255, 129}
\definecolor{limeAII}{RGB}{238, 255, 65}
\definecolor{limeAIV}{RGB}{198, 255, 0}
\definecolor{limeAVII}{RGB}{174, 234, 0}

\definecolor{yellowOV}{RGB}{255, 253, 231}
\definecolor{yellowI}{RGB}{255, 249, 196}
\definecolor{yellowII}{RGB}{255, 245, 157}
\definecolor{yellowIII}{RGB}{255, 241, 118}
\definecolor{yellowIV}{RGB}{255, 238, 88}
\definecolor{yellowV}{RGB}{255, 235, 59}
\definecolor{yellowVI}{RGB}{253, 216, 53}
\definecolor{yellowVII}{RGB}{251, 192, 45}
\definecolor{yellowVIII}{RGB}{249, 168, 37}
\definecolor{yellowIX}{RGB}{245, 127, 23}
\definecolor{yellowAI}{RGB}{255, 255, 141}
\definecolor{yellowAII}{RGB}{255, 255, 0}
\definecolor{yellowAIV}{RGB}{255, 234, 0}
\definecolor{yellowAVII}{RGB}{255, 214, 0}

\definecolor{amberOV}{RGB}{255, 248, 225}
\definecolor{amberI}{RGB}{255, 236, 179}
\definecolor{amberII}{RGB}{255, 224, 130}
\definecolor{amberIII}{RGB}{255, 213, 79}
\definecolor{amberIV}{RGB}{255, 202, 40}
\definecolor{amberV}{RGB}{255, 193, 7}
\definecolor{amberVI}{RGB}{255, 179, 0}
\definecolor{amberVII}{RGB}{255, 160, 0}
\definecolor{amberVIII}{RGB}{255, 143, 0}
\definecolor{amberIX}{RGB}{255, 111, 0}
\definecolor{amberAI}{RGB}{255, 229, 127}
\definecolor{amberAII}{RGB}{255, 215, 64}
\definecolor{amberAIV}{RGB}{255, 196, 0}
\definecolor{amberAVII}{RGB}{255, 171, 0}

\definecolor{orangeOV}{RGB}{255, 243, 224}
\definecolor{orangeI}{RGB}{255, 224, 178}
\definecolor{orangeII}{RGB}{255, 204, 128}
\definecolor{orangeIII}{RGB}{255, 183, 77}
\definecolor{orangeIV}{RGB}{255, 167, 38}
\definecolor{orangeV}{RGB}{255, 152, 0}
\definecolor{orangeVI}{RGB}{251, 140, 0}
\definecolor{orangeVII}{RGB}{245, 124, 0}
\definecolor{orangeVIII}{RGB}{239, 108, 0}
\definecolor{orangeIX}{RGB}{230, 81, 0}
\definecolor{orangeAI}{RGB}{255, 209, 128}
\definecolor{orangeAII}{RGB}{255, 171, 64}
\definecolor{orangeAIV}{RGB}{255, 145, 0}
\definecolor{orangeAVII}{RGB}{255, 109, 0}

\definecolor{deeporangeOV}{RGB}{251, 233, 231}
\definecolor{deeporangeI}{RGB}{255, 204, 188}
\definecolor{deeporangeII}{RGB}{255, 171, 145}
\definecolor{deeporangeIII}{RGB}{255, 138, 101}
\definecolor{deeporangeIV}{RGB}{255, 112, 67}
\definecolor{deeporangeV}{RGB}{255, 87, 34}
\definecolor{deeporangeVI}{RGB}{244, 81, 30}
\definecolor{deeporangeVII}{RGB}{230, 74, 25}
\definecolor{deeporangeVIII}{RGB}{216, 67, 21}
\definecolor{deeporangeIX}{RGB}{191, 54, 12}
\definecolor{deeporangeAI}{RGB}{255, 158, 128}
\definecolor{deeporangeAII}{RGB}{255, 110, 64}
\definecolor{deeporangeAIV}{RGB}{255, 61, 0}
\definecolor{deeporangeAVII}{RGB}{221, 44, 0}

\definecolor{brownOV}{RGB}{239, 235, 233}
\definecolor{brownI}{RGB}{215, 204, 200}
\definecolor{brownII}{RGB}{188, 170, 164}
\definecolor{brownIII}{RGB}{161, 136, 127}
\definecolor{brownIV}{RGB}{141, 110, 99}
\definecolor{brownV}{RGB}{121, 85, 72}
\definecolor{brownVI}{RGB}{109, 76, 65}
\definecolor{brownVII}{RGB}{93, 64, 55}
\definecolor{brownVIII}{RGB}{78, 52, 46}
\definecolor{brownIX}{RGB}{62, 39, 35}

\definecolor{grayOV}{RGB}{250, 250, 250}
\definecolor{grayI}{RGB}{245, 245, 245}
\definecolor{grayII}{RGB}{238, 238, 238}
\definecolor{grayIII}{RGB}{224, 224, 224}
\definecolor{grayIV}{RGB}{189, 189, 189}
\definecolor{grayV}{RGB}{158, 158, 158}
\definecolor{grayVI}{RGB}{117, 117, 117}
\definecolor{grayVII}{RGB}{97, 97, 97}
\definecolor{grayVIII}{RGB}{66, 66, 66}
\definecolor{grayIX}{RGB}{33, 33, 33}

\definecolor{bluegrayOV}{RGB}{236, 239, 241}
\definecolor{bluegrayI}{RGB}{207, 216, 220}
\definecolor{bluegrayII}{RGB}{176, 190, 197}
\definecolor{bluegrayIII}{RGB}{144, 164, 174}
\definecolor{bluegrayIV}{RGB}{120, 144, 156}
\definecolor{bluegrayV}{RGB}{96, 125, 139}
\definecolor{bluegrayVI}{RGB}{84, 110, 122}
\definecolor{bluegrayVII}{RGB}{69, 90, 100}
\definecolor{bluegrayVIII}{RGB}{55, 71, 79}
\definecolor{bluegrayIX}{RGB}{38, 50, 56}
\definecolor{bluegrayX}{RGB}{17, 23, 26}

\definecolor{myACMBlue}{cmyk}{1,0.1,0,0.1}
\definecolor{myACMYellow}{cmyk}{0,0.16,1,0}
\definecolor{myACMOrange}{cmyk}{0,0.42,1,0.01}
\definecolor{myACMRed}{cmyk}{0,0.90,0.86,0}
\definecolor{myACMLightBlue}{cmyk}{0.49,0.01,0,0}
\definecolor{myACMGreen}{cmyk}{0.20,0,1,0.19}
\definecolor{myACMPurple}{cmyk}{0.55,1,0,0.15}
\definecolor{myACMDarkBlue}{cmyk}{1,0.58,0,0.21}

\newcommand{\link}[1]{{\href{#1}{\color{blueVI}\textbf{\texttt{#1}}}}}

\crefname{figure}{fig.}{fig.}
\Crefname{figure}{Fig.}{Fig.}
\crefname{equation}{eq.}{eq.}
\Crefname{equation}{Eq.}{Eq.}
\crefname{section}{\S}{\S}
\creflabelformat{equation}{#2\textup{#1}#3}

\definecolor{myReferenceURLColor}{HTML}{09326F}
\newcommand{\figpart}[1]{\textcolor{blueVIII}{#1}}
\hypersetup{
  colorlinks,
  linkcolor=blueVIII,
  citecolor=blueVIII,
  urlcolor=myReferenceURLColor,
  filecolor=myACMDarkBlue
}

\newcommand{\tool}{\textsc{NOVA}}

\definecolor{soulorange}{RGB}{255, 212, 153}
\definecolor{soulgray}{RGB}{220, 220, 220}
\definecolor{soulgraylight}{RGB}{235, 235, 235}
\definecolor{soulred}{RGB}{252, 217, 218}
\definecolor{soulbluelight}{RGB}{208, 233, 253}
\definecolor{souldorangelight}{RGB}{254, 234, 212}
\colorlet{soulblue}{blueV!30}

\definecolor{tagbordercolor}{rgb}{0.8, 0.8, 0.8}
\definecolor{tagbgcolor}{rgb}{0.9, 0.9, 0.9}

\newtcbox{\tagg}{nobeforeafter, colframe=tagbordercolor,
colback=tagbgcolor, boxrule=0.5pt, arc=1pt,
  boxsep=0pt,left=2pt,right=2pt,top=1.5pt,bottom=2pt,tcbox raise base}

\definecolor{lightgray}{RGB}{247, 247, 247}
\definecolor{midgray}{RGB}{179, 179, 179}

\newtcbox{\featuretag}{on line,
  colframe=midgray,colback=lightgray,
  boxrule=0.5pt,arc=2pt,boxsep=0pt,left=2pt,right=1pt,top=1pt,bottom=1pt
}
\definecolor{tagbgcolor}{rgb}{1, 1, 1}

\definecolor{boxyellow}{RGB}{206, 171, 1}
\definecolor{boxgreen}{RGB}{14, 152, 136}
\definecolor{boxblue}{RGB}{77, 167, 223}

\setul{0.45ex}{0.2ex}

\begin{document}

\firstsection{Introduction}

\maketitle

Our VIS community has developed a large number of visual analytics (VA) tools that can help data scientists in a wide range of tasks,
from exploring and comparing data~\cite{heimerlEmbCompVisualInteractive2020}, conducting analytical experiments~\cite{wangVisualCausalityAnalyst2016}, to interpreting machine learning (ML) models~\cite{noriInterpretMLUnifiedFramework2019}.
Many VA tools are web applications that users access through their web browsers~(\autoref{fig:teaser}\figpart{A}).
However, it can be challenging for data scientists to adopt these systems, as they require data scientists to tailor data structure to fit different systems, switch between multiple programming environments, and develop new workflows to streamline VA processes.
Thus, Bertini~\cite{bertiniBuildingEasyToAdoptSoftware2022} suggests that, to make a greater impact, we should build VA tools that can be easily adopted.
His viewpoint resonates with many VIS researchers on \href{https://twitter.com/filwd/status/1514654384990400514}{\color{blueVI}\textbf{\texttt{social media}}}.

On the other hand, computational notebooks such as Jupyter Notebook~\cite{kluyverJupyterNotebooksPublishing2016} have become the most popular programming environment among data scientists~\cite{kaggleStateMachineLearning2021}.
These notebooks present a fluid programming environment that consists of any number of cells---small editors for markdown text and code.
Users can execute a code cell, and then its output, such as texts, tables, and visualizations, will be displayed below the cell.
Visualizations play a critical role in the success of notebooks:
data scientists appreciate that they can rapidly visualize dataset and machine learning models while exploring the dataset and building ML models~\cite{ruleExplorationExplanationComputational2018}.
Therefore, to develop VA tools that can be easily integrated into existing data science workflows, visualization researchers and practitioners can enhance their tools with computational notebook compatibility~(\autoref{fig:teaser}\figpart{B}).
To lower the barrier to developing notebook-ready VA tools, we \textbf{contribute}:

\begin{itemize}[topsep=1pt, itemsep=0mm, parsep=3pt, leftmargin=9pt]

  \item \textbf{\tool{}, a simple and flexible method} to adapt existing web-based VA tools into computational notebook widgets.
  End-users can easily install these widgets and use them directly in notebooks.
  NOVA works with all popular web development technologies, such as Svelte, React, and Vanilla JS.
  The result widgets support all types of popular computational notebooks, such as Jupyter Notebook/Lab, Google Colab, and VSCode Notebook.
  With NOVA, VA researchers have developed widgets that help data scientists easily curate~\cite{wangTimberTrekExploringCurating2022a}, audit~\cite{munechikaVisualAuditorInteractive2022a}, and edit~\cite{wangGAMChangerEditing2021} ML models.

  \item \textbf{Detailed examples of applying \tool{}} to convert a graph VA tool to a notebook widget that users can easily access in notebooks~(\autoref{fig:teaser}).
  We present three implementations of this graph VA tool with Svelte, React, and Vanilla JS, and demonstrate how to apply \tool{} on each implementation.
  We also show how one can create a web page that demonstrates a VA tool in both web app and notebook widget forms simultaneously.
  To continuously improve \tool{} and provide up-to-date examples, we maintain a list of VA tools that apply \tool{}.
  This list and all examples can be accessed at \link{https://github.com/poloclub/nova}.

\end{itemize}

\noindent We hope our work can inspire and inform future design, research, and development of VA tools that fit into data scientists' workflows. %

\section{\tool{} Method}

\tool{} supports any web-based VA tools that do not require a dedicated backend server: many existing VA tools run entirely in the browser without a backend~\cite[e.g.][]{wangGAMChangerEditing2021,wangTimberTrekExploringCurating2022a,munechikaVisualAuditorInteractive2022a}.
NOVA creates a Python package that launches the VA interface with end-user specified data in a notebook cell.
In this section, we outline the three simple steps of applying \tool{}.
Each step’s implementation details vary depending on the VA tool’s development stack; readers can refer to three detailed examples with different development stacks on \href{https://github.com/poloclub/nova}{\color{blueVI}\textbf{\texttt{GitHub}}}.\looseness=-1

\begin{enumerate}[topsep=2pt, itemsep=0mm, parsep=3pt, leftmargin=17pt, label=\textbf{S\arabic*.}, ref=S\arabic*]
  \item \label{item:s1}
  \textbf{Convert the VA tool into a single HTML file.}
  We first combine a VA tool's mark-ups, style sheets, scripts, and assets into a single HTML file.
  This allows us to embed the VA tool as a configurable \texttt{iframe} widget in notebooks.
  This step is simple: with module bundlers, such as \texttt{Rollup} and \texttt{Webpack}, one can bundle their VA tool into an HTML file with one command.

  \item \label{item:s2}
  \textbf{Design Python wrapper API.}
  To enable end-users to feed data and configurations into the VA widget, we can create a Python function API that (1) displays the widget in a notebook cell, and (2) collects and sends user input to the widget through standard Web Events.
  Thus, the VA widget can listen to these events and behave accordingly.
  This one-directional messaging approach is generalizable to all types of web-based notebooks.

  \item \label{item:s3}
  \textbf{Publish the VA widget in a software repository.}
  Finally, we package the VA widget as a Python library and publish it in the software repository Python Package Index (PyPI).
  This enables end-users to install the widget with one command and access it in their notebooks.
  Our examples use Python, as it is the most popular programming language among data scientists~\cite{kaggleStateMachineLearning2021}.
  However, one can easily generalize \tool{} to other languages, such as R and Julia, and publish the widgets in corresponding software repositories, such as CRAN and Julia Registries.
  As publishing libraries in software repositories is a well-established practice, one can usually complete this step in 10 minutes.
\end{enumerate}
Many VA tools provide companion demo pages that users can easily access and try out the tool~\cite[e.g.][]{wangGAMChangerEditing2021,munechikaVisualAuditorInteractive2022a}.
Leveraging recent advancements in Python's in-browser programming environment, \tool{} provides instruction on creating a web page that demonstrates a VA tool in both web app form and notebook widget form simultaneously.\looseness=-1
\section{\tool{} in Action}

\tool{} has already been used in real-life VA applications.
A recent example is \textsc{GAM Changer}~\cite{wangGAMChangerEditing2021}---a VA tool that enables data scientists and domain experts to edit model weights of Generalized Additive Models (GAMs).
\textsc{GAM Changer} is originally developed as a stand-alone web app using Svelte, JavaScript, and WebAssembly; we then apply \tool{} to create a notebook widget.
We first bundle this tool with all assets including images and WebAssembly binaries into an HTML file with \texttt{Rollup} bundler~(\ref{item:s1}).
Then, we design a simple Python function API that allows end-users to feed \textsc{GAM Changer} with their data and model weights and control the widget's size~(\ref{item:s2}).
Finally, we publish the notebook widget on PyPI~(\ref{item:s3}) and open source it under the \texttt{InterpretML} ecosystem~\cite{noriInterpretMLUnifiedFramework2019}.
Feedback from data scientists and physicians highlights that \textsc{GAM Changer’s} notebook widget fits into their ML development workflows, and they appreciate using this tool directly in notebooks.

Interestingly, VA notebook widgets not only enhance data scientists’ analytical workflow, but a notebook environment also improves the user experience of VA tools themselves.
In \textsc{GAM Changer}, users can only inspect and edit the model’s behavior on one feature at a time; users sometimes would like to compare the model's prediction patterns across multiple features simultaneously.
To do that with the web app, users would need to follow a cumbersome process to open the app in multiple browser windows and then resize and arrange these windows.
In contrast, with \textsc{GAM Changer’s} notebook widget, users can simply launch this tool in multiple cells in the same notebook.
Moreover, with sticky notebook cells~\cite{wangStickyLandBreakingLinear2022}, users can even arrange these notebook cells in non-linear ways and mix-and-match multiple VA tools to create a full-fledged dashboard.
\vspace{-13pt}
\section{Discussion \& Future Work}
\vspace{-1pt}

To support diverse VA development technologies and different types of computational notebooks, \tool{} has made two compromises.
First, \tool{} only supports web-based VA tools that do not require a dedicated backend server or tools whose backend code can be replaced with front-end scripts.
It is possible to host a web server inside a Jupyter Notebook and launch a VA widget with a backend~\cite{wexlerWhatIfToolInteractive2019}, however, this functionality is not supported by all computational notebooks.
Also, \tool{} only supports one-way communication from the notebook to the VA tool.
With \href{https://github.com/jupyter-widgets/ipywidgets}{\color{blueVI}\textbf{\texttt{ipywidgets}}}, it is possible to create a VA widget that supports two-way communication~\cite{cabreraCreatingReactiveJupyter2020}.
However, for security reasons, other notebooks besides Jupyter Notebook and Lab do not support \texttt{ipywidgets} or similar two-way communication protocols.
Many existing VA tools do not require a backend server nor two-way communications~\cite[e.g.][]{wangGAMChangerEditing2021,wangTimberTrekExploringCurating2022a,munechikaVisualAuditorInteractive2022a}.\looseness=-1

We maintain a public list of open-source VA tools that apply \tool{} to create notebook widgets, providing up-to-date examples for VIS researchers and practitioners.
It also helps us monitor VA development trends and continuously improve this workflow.
For future work, we will reflect on our experience with \tool{} and engage with diverse stakeholders to distill guidelines and best practices for designing and developing in-notebook VA tools.

\textbf{Conclusion.}
We present \tool{}, a simple and flexible method to convert existing web-based VA tools to notebook widgets that can be easily accessed in computational notebooks.
Three detailed examples with a toy VA tool demonstrate how one can apply \tool{} on VA tools developed with different web technologies.
We hope our work will lower the barrier to creating interactive visualization systems that work in computational notebooks and inspire future researchers to create VA tools that fit into data scientists' workflows. %

\vspace{-4pt}
\acknowledgments{
  This work was supported in part by a J.P. Morgan PhD Fellowship, as well as gifts from Cisco and Bosch.
}

\bibliographystyle{abbrv-hyperref}

\setlength{\bibsep}{0pt}
{\footnotesize
\bibliography{23-nova}

\begin{thebibliography}{10}

\bibitem{bertiniBuildingEasyToAdoptSoftware2022}
E.~Bertini.
\newblock
  \href{https://filwd.substack.com/p/building-easy-to-adopt-software-while}{Building
  ({{Easy-To-Adopt}}) {{Software}} while {{Doing Visualization Research}}},
  2022.

\bibitem{cabreraCreatingReactiveJupyter2020}
{\'A}.~A. Cabrera.
\newblock
  \href{https://cabreraalex.medium.com/creating-reactive-jupyter-widgets-with-svelte-ef2fb580c05}{Creating
  {{Reactive Jupyter Widgets With Svelte}}}, 2020.

\bibitem{heimerlEmbCompVisualInteractive2020}
F.~Heimerl, C.~Kralj, T.~Moller, and M.~Gleicher.
\newblock \href{https://doi.org/10.1109/TVCG.2020.3045918}{{{embComp}}:
  {{Visual Interactive Comparison}} of {{Vector Embeddings}}}.
\newblock {\em IEEE TVCG}, 2020.

\bibitem{kaggleStateMachineLearning2021}
Kaggle.
\newblock \href{https://www.kaggle.com/kaggle-survey-2021}{State of {{Machine
  Learning}} and {{Data Science}} 2021}, 2021.

\bibitem{kluyverJupyterNotebooksPublishing2016}
T.~Kluyver and {others}.
\newblock \href{https://doi.org/10.3233/978-1-61499-649-1-87}{Jupyter
  {{Notebooks}} - a {{Publishing Format}} for {{Reproducible Computational
  Workflows}}}.
\newblock {\em ELPUB}, 2016.

\bibitem{munechikaVisualAuditorInteractive2022a}
D.~Munechika, Z.~J. Wang, J.~Reidy, J.~Rubin, K.~Gade, K.~Kenthapadi, and D.~H.
  Chau.
\newblock \href{https://doi.org/10.1109/VIS54862.2022.00018}{Visual
  {{Auditor}}: {{Interactive Visualization}} for {{Detection}} and
  {{Summarization}} of {{Model Biases}}}.
\newblock {\em IEEE VIS}, 2022.

\bibitem{noriInterpretMLUnifiedFramework2019}
H.~Nori, S.~Jenkins, P.~Koch, and R.~Caruana.
\newblock \href{http://arxiv.org/abs/1909.09223}{{{InterpretML}}: {{A Unified
  Framework}} for {{Machine Learning Interpretability}}}.
\newblock {\em arXiv}, 2019.

\bibitem{ruleExplorationExplanationComputational2018}
A.~Rule, A.~Tabard, and J.~D. Hollan.
\newblock \href{https://doi.org/10.1145/3173574.3173606}{Exploration and
  {{Explanation}} in {{Computational Notebooks}}}.
\newblock In {\em {{CHI}}}, 2018.

\bibitem{wangVisualCausalityAnalyst2016}
J.~Wang and K.~Mueller.
\newblock \href{https://doi.org/10.1109/TVCG.2015.2467931}{The {{Visual
  Causality Analyst}}: {{An Interactive Interface}} for {{Causal Reasoning}}}.
\newblock {\em IEEE TVCG}, 22, 2016.

\bibitem{wangStickyLandBreakingLinear2022}
Z.~J. Wang, K.~Dai, and W.~K. Edwards.
\newblock \href{https://doi.org/10.1145/3491101.3519653}{{{StickyLand}}:
  {{Breaking}} the {{Linear Presentation}} of {{Computational Notebooks}}}.
\newblock {\em CHI EA}, 2022.

\bibitem{wangGAMChangerEditing2021}
Z.~J. Wang, A.~Kale, H.~Nori, P.~Stella, M.~Nunnally, D.~H. Chau,
  M.~Vorvoreanu, J.~W. Vaughan, and R.~Caruana.
\newblock \href{http://arxiv.org/abs/2112.03245}{{{GAM Changer}}: {{Editing
  Generalized Additive Models}} with {{Interactive Visualization}}}.
\newblock {\em arXiv:2112.03245}, 2021.

\bibitem{wangTimberTrekExploringCurating2022a}
Z.~J. Wang, C.~Zhong, R.~Xin, T.~Takagi, Z.~Chen, D.~H. Chau, C.~Rudin, and
  M.~Seltzer.
\newblock \href{https://doi.org/10.1109/VIS54862.2022.00021}{{{TimberTrek}}:
  {{Exploring}} and {{Curating Sparse Decision Trees}} with {{Interactive
  Visualization}}}.
\newblock In {\em {{VIS}}}, 2022.

\bibitem{wexlerWhatIfToolInteractive2019}
J.~Wexler, M.~Pushkarna, T.~Bolukbasi, M.~Wattenberg, F.~Viegas, and J.~Wilson.
\newblock \href{https://doi.org/10.1109/TVCG.2019.2934619}{The {{What-If
  Tool}}: {{Interactive Probing}} of {{Machine Learning Models}}}.
\newblock {\em TVCG}, 26, 2019.

\end{thebibliography}
}

\end{document}